\documentclass[runningheads]{llncs}

\usepackage[utf8]{inputenc}
\usepackage{fullpage}
\usepackage{amsmath,amssymb}
\usepackage{graphicx,subcaption}
\graphicspath{ {./images/} }
\usepackage[capitalize]{cleveref}
\usepackage{enumitem}

\begin{document}
\title{Fast Agent-Based Simulation Framework with Applications to Reinforcement Learning and the Study of Trading Latency Effects}
\titlerunning{MAXE}
%
\author{Peter Belcak\inst{1,2} \and
Jan-Peter Calliess\inst{1,3} \and
Stefan Zohren\inst{1,4}}
\authorrunning{P. Belcak et al.}
%
\institute{Oxford-Man Institute of Quantitative Finance, University of Oxford, Oxford, UK
\and \email{peter.belcak@st-hildas.ox.ac.uk} \and  \email{jan-peter.calliess@oxford-man.ox.ac.uk} \and \email{stefan.zohren@eng.ox.ac.uk} \\
}


\maketitle              


\begin{abstract}
We introduce a new software toolbox for agent-based simulation. Facilitating rapid prototyping by offering a user-friendly Python API, its core rests on an efficient C++ implementation to support simulation of large-scale multi-agent systems.
Our software environment benefits from a versatile message-driven architecture. Originally developed to support research on financial markets, it offers the flexibility to simulate a wide-range of different (easily customisable) market rules and to study the effect of auxiliary factors, such as delays, on the market dynamics. As a simple illustration, we employ our toolbox to investigate the role of the order processing delay in normal trading and for the scenario of a significant price change.

Owing to its general architecture, our toolbox can also be employed as a generic multi-agent system simulator. We provide an example of such a non-financial application by simulating a mechanism for the coordination of no-regret learning agents in a multi-agent network routing scenario previously proposed in the literature. 


\end{abstract}


\newcommand{\ffrac}[2]{\ensuremath{\frac{\displaystyle #1}{\displaystyle #2}}}
\newcommand{\jcom}[1]{\textcolor{cyan} {\emph{Jan says: #1}}}
\newcommand{\pcom}[1]{\textcolor{red} {\emph{Peter says: #1}}}
\newcommand{\tdcom}[1]{\textcolor{magenta} {\emph{TODO: #1}}}


\section{Motivation}
Complementing the classical methods of statistical analysis and mathematical modelling, 
agent-based modelling (ABM) of financial markets has recently been gaining traction  \cite{luna2012economic,iori2012agent,cont2007volatility,buchanan2009meltdown}.
In particular, applications of this paradigm to market microstructure \cite{bouchaud2018trades} have attracted increasing attention.
To name but a few, they include the study of statistical properties of limit order books \cite{bouchaud2002statistical}, (non-)strategic behavior of a collective of traders \cite{farmer2006random} when modelled via the flow of their orders, as well as research into market bubbles and crashes \cite{paddrik2012agent}.
With the ever-increasing importance of automated trading in finance and the rising popularity of artificial intelligence in academic and industrial research, the importance of the ABM approach in the study of electronic markets is likely to grow further.

The diversity of use cases of ABM in finance and economics is reflected by the recent proliferation of a variety of software tools tailored to the particularities of their respective applications, as can be seen in the aforementioned sources. 
What is missing is an efficient code base implementing a general, all-encompassing multi-agent exchange framework that can be easily adapted to simulate scalable ABMs based on any particular exchange as a special case. 
Among many other conceivable use cases, such a software environment could serve as a flexible toolbox allowing its users to investigate a range of research questions. Such could include, but are not limited to, the following:
\begin{itemize}
  \item The impact of different matching algorithms on the (learned) behaviour and revenues of (adaptive) trader agents inhabiting a given limit order book (LOB);
  \item The amount of strategic decision making required to explain some of the important statistical properties of these LOBs;
  \item The response of strategic trader agent behavior to a change in the rules of the order matching, as well as to changing infrastructural effects such as communication delays.
  \item Conversely to the above, the impact of different learning behaviors of the trading agents on the ensuing market dynamics.
\end{itemize}

To address the need for such a toolbox, we introduce the Multi-Agent eXchange Environment (MAXE), a general code environment for the simulation of agent-based models, with a database ready-to-use agents for simulation of electronic exchanges and other financial markets. For convenience, MAXE also provides a Python API to facilitate rapid prototyping of artificial agents. However, since the meaningfulness of ABMs often rests on the capability to simulate large agent populations, the core of the implementation was written in C++, with an eye for computational and memory efficiency, as well as for support for native multi-threading for execution of separate simulation instances (with possibly varying parameters) in parallel.

The remainder of this paper is structured as follows: After placing our toolbox into the context of previous, related simulator packages in Sec. \ref {sec:related_work}, Sec. \ref {sec:architecture} proceeds with introducing the architecture of MAXE. 
We present different use cases of our framework. Sec. \ref{sec:delay} contains an illustration of a simple study of the effects of communication delays. Sec. \ref{sec:example_use} shows how MAXE can be utilised in the general context of agent-based modelling, and Sec. \ref{sec:performance_comparison} compares MAXE's performance in a simple simulation scenario to that of a contender. Concluding remarks can be found in Sec. \ref {sec:conclusions}.


\section{Related Work} \label{sec:related_work}
Beyond simple market replay approaches, there still is a need for publicly available ABM software sufficiently generic to be capable of simulating the markets and many other environments at scale. Our toolbox was designed to meet this demand. The most closely related toolboxes we are aware of include Adaptive Modeler \cite{adaptiveModeller}, Swarm \cite{swarmSoftware}, NetLogo \cite{sklar2007netlogo}, Repast \cite{collier2003repast}, and ABIDES \cite{byrd2019abides}.
In what is to follow, we briefly summarise the features of these packages in relation to ours.

\emph{Adaptive Modeler} \cite{adaptiveModeller} is a ``freemium'' specialized market simulator first released in 2003 and still maintained. At the core of the software is a virtual market featuring a predefined set of classes of agents that may be further adjusted by the user by changing various parameters such as the population sizes, agent wealth, or class mutation probability. Once an environment consisting of traders and traded assets is specified, the user may start the simulation whilst keeping track of outputs such as the event log, quotes, or various economic statistics. All of these functionalities are -- or can easily be -- implemented in MAXE as well. In addition however, MAXE also allows the creation of completely customised agents with arbitrary behavior and simulate them on an arbitrary timescale, as the unit time step is not bound to any physical measure of time and can thus be chosen to represent an arbitrarily small fraction of a second.

\emph{Swarm} \cite{swarmSoftware} is an open-source ABM package for simulating the interaction of agents and their emergent collective behaviour. First released in 1999, it remains maintained today. Whilst not directly designed for financial modelling, it has been used to create the Santa Fe Artificial Stock Market \cite{palmer1999artificial} that, for the first time, reproduced a number of stylized facts about the behaviour of traders and further emphasized the importance of modelling of financial markets. Unlike swarm, MAXE comes with an incorporated time-tracking unit that takes care of the delivery of messages between the agents involved and the advancement of simulation time. This allows for a transparent unified channel of inter-agent communication, enabling simple scheduling of agent tasks (as outlined in an example in \cref{fig:communicationDiagram}) and greatly simplifying output generation and debugging.

\emph{NetLogo} \cite{sklar2007netlogo} and \textit{RePast} \cite{collier2003repast} are general-purpose software frameworks for agent-based modelling.
On top of simulation capabilities, both of these tools feature components enabling easy display of data, and have extensively been used for research in social sciences.
They have been previously used for small-scale simulation of financial markets, though their distributions do not feature readily available agents for market simulation.
In comparison to the extensive constraints placed on the agent interface by either, MAXE places no constraints on the design of the agent, apart from requiring that the agents conform to the minimal messaging structure, for which there are template agent classes readily available.

\emph{JADE} \cite{bergenti2020first} is an open-source software framework for peer-to-peer agent based applications.
It is fully implemented in Java, and as such suffers from the memory limitations and reliance on garbage collecting as it exists in a managed environment. 
It, however, benefits from the platform of choice by being runnable on every operating system supporting the Java runtime binaries, and even allows its users to run simulations across multiple devices simultaneously.

\emph{ABIDES} \cite{byrd2019abides} is the newest open source market modelling tool. Released as recently as 2019, it was specifically designed for LOB simulation. Aimed to closely resemble NASDAQ by implementing the NASDAQ ITCH and OUCH messaging protocols it hopes to offer itself as a tool for facilitation of AI research on the exchange. Just as MAXE, ABIDES and allows users to implement their own agents in Python. However, since MAXE also allows specification in C++ we expect that MAXE has an edge in terms of the execution efficiency and scalability. Moreover, MAXE, being based on a compiled binary core interacting directly with the operating system allows for multi-threaded execution of simulations, which becomes an advantage when simulating a range of similar simulations differing only in a number of input parameters. Apart from that, MAXE comes with the implicit support for the trading at multiple exchanges at once and for limit order books matched with different matching algorithms, in particular pro-rata matching. MAXE is also highly modular due to the option to develop a database of agents first, and then configure a set of simulations via an XML configuration file.


\section{Architecture} \label{sec:architecture}
MAXE is based on a message-driven, incremental protocol. Its core logic steps forward time and delivers messages, and thus while it was developed with the modelling of financial markets in sight, it can easily be utilised for simulating many general multi-agent systems unrelated to finance, an example of which we give in Sec. \ref{sec:example_use}. We will, nevertheless, continue to present MAXE's features mainly in the light of market simulation, believing that it will turn out to be its most popular application.

In MAXE, every relevant entity of a trading system one would wish to model (e.g. exchanges, traders, news outlets or social media) can be implemented as an agent. This is different to the usual approach to agent-based modelling of exchanges where at the centre of the simulation is the exchange concerned and the communication protocol between the entities of the trading system is made to resemble the one of the real exchange, often to ease the transition of any models developed there into production environments.
As it is the case with any common implementation of message-driven frameworks, agents taking part in the simulation remain dormant at any point in simulation time unless they have been delivered a message. 
When a message is due to be delivered, the simulation time freezes as all agents that have been delivered at least one message begin to take turns to deal with their inbox. Each agent is given an unlimited amount of execution time to process the messages they have been delivered and to send messages on their own. Messages can be dispatched either immediately (i.e. with zero delay) or scheduled to be delivered later in the future by specifying a non-negative delay which can be used to, for example, model latency, that we will show in Sec. \ref{sec:delay}.

\begin{figure}
    \centering\includegraphics[scale=0.20]{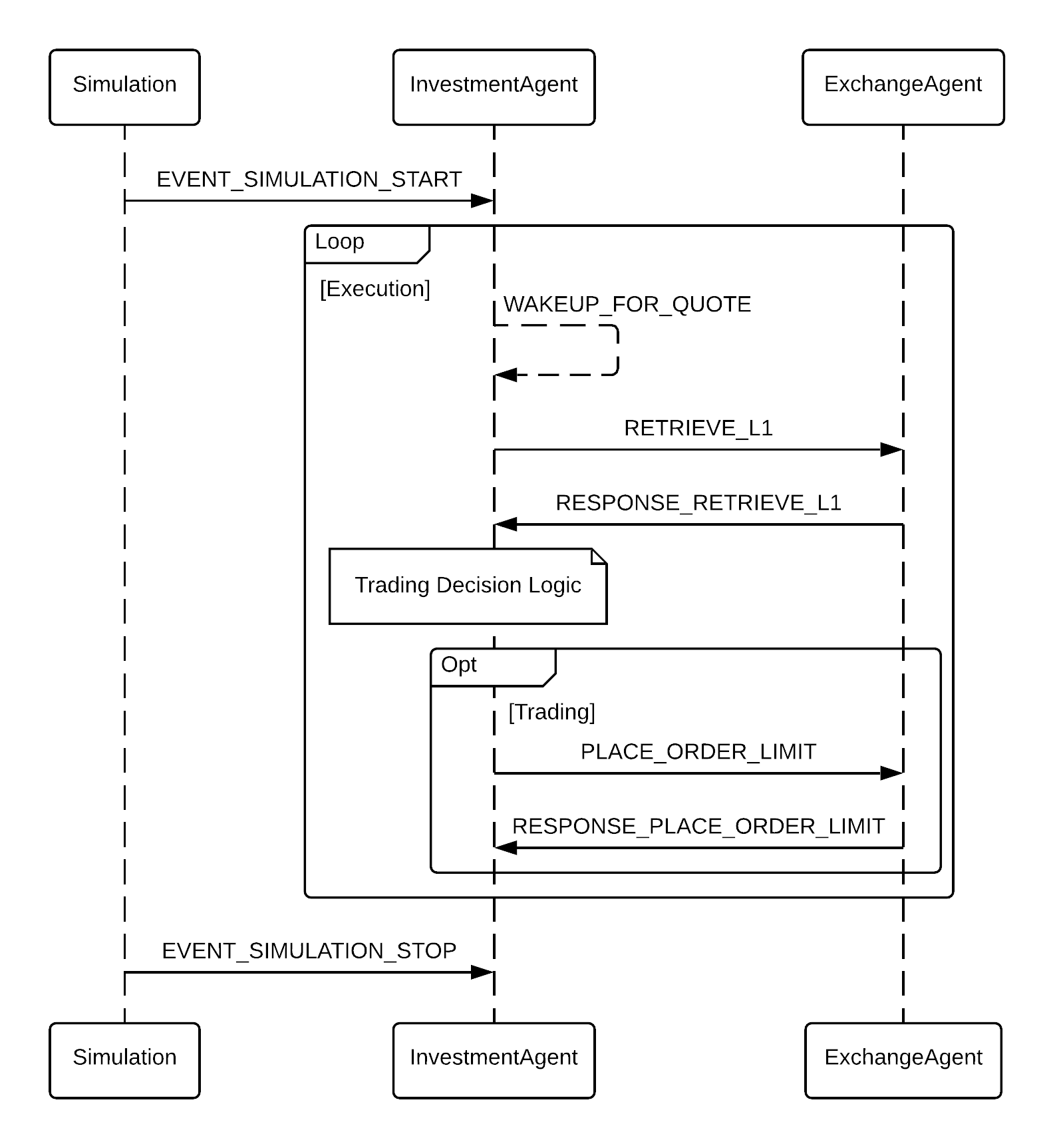}
    \caption{A sequence diagram of an example communication between a trading agent, exchange agent, and the simulation environment.}
    \label{fig:communicationDiagram}
\end{figure}

At the beginning of a simulation, each agent is delivered a message that allows them to take initial actions and possibly schedule a wake-up in the future by addressing a message to themselves. At the end of the simulation, a message of similar nature is sent out to all agents to allow them to process and save any data they might have been gathering up to that point for further analysis outside the simulation environment. \cref{fig:communicationDiagram} shows an example communication of an agent that trades based on regular L1 quote data from the exchange. 

\begin{figure*}
    \centering\includegraphics[scale=0.20]{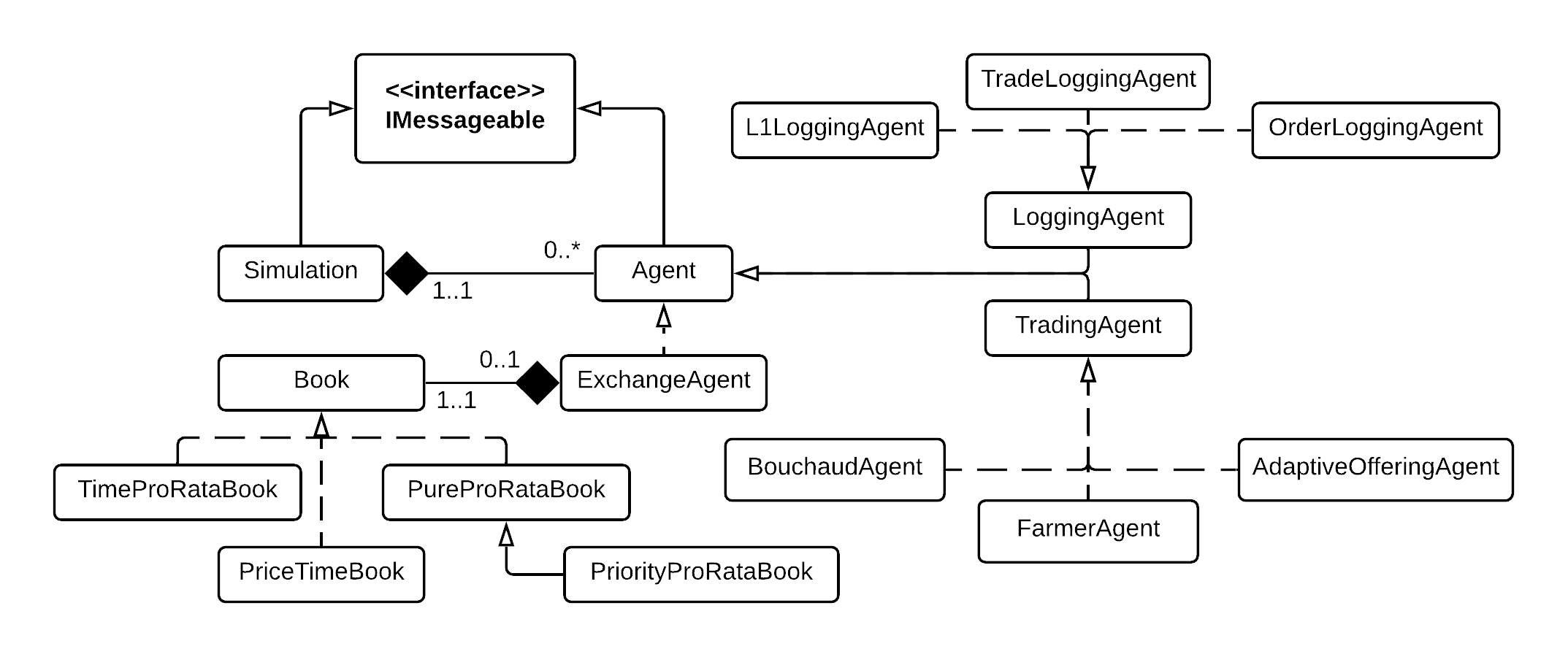}
    \caption{
        The class diagram of the Simulation-Agent hierarchy of the simulator.
    }
    \label{fig:classDiagram}
\end{figure*}

Aside from its core, MAXE also contains a small initial repository of common agents that can be expanded upon by its users. This initial repository includes an exchange agent that can operate a number of different matching mechanism, as well as a collection of zero-intelligence and other simple agents. An overview of the top level of the hierarchy of available agents is depicted in \cref{fig:classDiagram}. Further details on the various agents can be found in the code repository \cite{maxeGitHub}.


For simplicity and in order to facilitate convenient prototyping of trading system models, MAXE has been built with an interactive console interface, designed to read the simulation configurations from a hand-editable XML file, and is, despite the overall emphasis on the performance, supplemented with an additional Python interface.
A user of MAXE wishing to quickly try out an idea for an agent-based model would thus proceed as follows: First, they would consider whether any of the built-in agent types fit their needs. For any agent type with custom behaviour they would write a Python script, testing it in a `mock' environment API that is provided. Once satisfied with the scripted behaviour of individual agents, they would with ease set up the simulation of their model by writing an XML configuration file, and once satisfied with the overall model, they would have the option to scale up to hundreds of thousands of agents by re-implementing the behaviour first scripted in Python in C++.
Some resource limits for MAXE when running solely agents implemented in C++ are discussed in Sec. \ref{sec:performance_comparison}.

\section{Example case study-- processing delay in market dynamics} \label{sec:delay}
As discussed previously, MAXE can serve as a simulation environment of many types of multi-agent systems. As a first example related to financial markets, 
we demonstrate MAXE's ability to simulate some aspects of ``market physics''. In particular, we utilise it to examine the effect processing or communication delays have on various statistics of the market dynamics following a large trade.

\subsection{The Model}
The core of the model consists of one exchange agent with a modifiable choice of matching algorithm and a population of zero-intelligence trading agents interacting with the exchange. The exchange agent maintains the limit order book and executes orders submitted by the trading agents. At the beginning of the simulation, the LOB contains two small orders, one on each side of the book with the initial bid-ask spread $S_0$, to serve as the indicators of the opening prices for further trading.

Following the start of the simulation, traders place orders and are given a fixed period of time to reconstruct the LOB to match the empirical average shape from \cite{bouchaud2002statistical} by placing orders in a manner described below. In the simulation runs focused on statistics not related to the study of impactful trades, the remaining time is used to measure those. The other type of simulations experiences an impact agent entering the exchange and making a large trade, following which more statistics are computed. The simulation runs over a fixed time horizon of $40000t_\text{p}$, chosen by experimentation focused on the setup appearing to have dealt with the largest of the trades used in our experiment.

The behaviour of our trading agent is similar to the behaviour presented in \cite{bouchaud2002statistical} that has been previously shown to be able to reconstruct the LOB's shape to be resembling the one of real LOBs of highly liquid stocks on the Paris Bourse. The behavior presented in \cite{bouchaud2002statistical} is further adjusted by some features of the behaviour presented in \cite{farmer2006random}, which has been shown to be able to explain some of the dynamic properties of the LOB, including the variance of the bid-ask spread. For a detailed specification of the agents' behaviour we refer the reader to \cite{maxeGitHub}.

According to the L1 information available to the trader at the time they are making the decision (which may be outdated due to the communication delay between the trader and the exchange), each trader places both market and not immediately marketable limit orders according to a Poisson process with rate $r_{\text{p}}$, with the fraction of the market orders $f_{\text{m}}$ being a parameter of the simulation.

Each order has lifetime distributed according to the exponential distribution with mean $t_{\text{l}}$ that was a parameter of the simulation. Thus, the stream of the cancellation orders can be thought of as a marked Poisson process with rate $r_{\text{c}} = \frac{1}{l_{\text{o}}}$ and where the marking probability is inversely proportional to the number of orders in the LOB. The price of the order is drawn from the empirical power-law distribution relative to the best price at the time of observation.

We define the \emph{processing delay} $d$, or simply \emph{delay} to be the duration between the time the information about the state of the limit order book is produced for trading agents and the time when the exchange processes the agent's order against the LOB. This time includes the two-way latency between the agent and the exchange, the (simulation) time it takes the exchange to process the queue of incoming orders, and decision time on the trader's side. Furthermore, taking the zero-intelligence approach to model the trading and the limit order book as a whole, the processing delay can also be thought of as encapsulating the time it takes the trader to decide whether and how to trade and possibly evaluating their strategy given the information becoming available during that time, and we we shall use this fact when interpreting our findings. We also define \emph{greed} $g$ to be the size of a large market order expressed as a fraction of the total volume (i.e., considering the volume of all price levels) in the queue it is meant to be executed against.

\subsection{Findings}
When simulating, we treated the placement frequency $r_{\text{p}}$ as fixed and looked at the effects of the other two time-based parameters, $r_{\text{c}}$ and $d$, relative to it. The observed effects turned out to be independent of the matching algorithm used. Perhaps somewhat more surprisingly, the cancellation rate $r_{\text{c}}$ appeared not to have had any effect on the statistics considered (see below).

\textbf{Notation: } If $Q$ is the quantity we are observing, let $e[Q]$ denote the empirical simulation-time-weighted mean of $Q$ and $v[Q]$ the empirical simulation-time-weighted variance of $Q$.

\begin{figure*}
    \centering

    \captionsetup[subfigure]{justification=centering}
    \begin{subfigure}[t]{.3\linewidth}
        \centering\hspace{-0.8cm}\includegraphics[width=1.0\linewidth, height=1.0\linewidth]{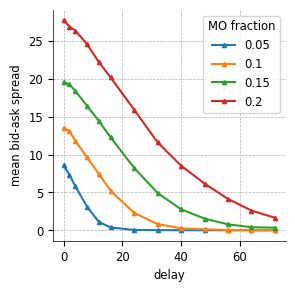}
        \caption{}
        \label{fig:bidAskFigure}
    \end{subfigure}
    \begin{subfigure}[t]{.3\linewidth}
        \centering\hspace{-0.85cm}\includegraphics[width=1.0\linewidth, height=1.0\linewidth]{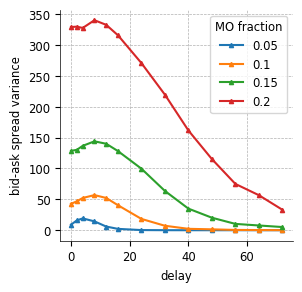}
        \caption{}
        \label{fig:varianceBidAskFigure}
    \end{subfigure}
    \begin{subfigure}[t]{.3\linewidth}
        \centering\hspace{-0.85cm}\includegraphics[width=1.0\linewidth, height=1.0\linewidth]{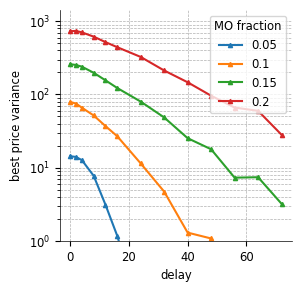}
        \caption{}
        \label{fig:varianceBestAskFigure}
    \end{subfigure}
\caption{Statistics of the simulation L1 data, namely the mean bid-ask spread, bid-ask spread variance, and the variance of the best price, plotted against $d$ (in multiples of $t_\text{p}$) for different values of $f_{\text{m}}$.}
\end{figure*}

We found that the mean bid-ask spread $e[S]$ increased linearly with the fraction of market orders $f_{\text{m}}$ (with a hint of convexity), decreased with $d$, and appeared to converge to the bid-ask spread of the initial setup $S_0$, coming within a few ticks distance of $S_0$ for all sufficiently large delays $d$. The relationship between the parameters involved is depicted in \cref{fig:bidAskFigure} and fitted ($R^2 = 0.90$)
$$
    e[S] \approx S_0 + s_0 f_m e^{-s_1 d}.
$$

The time-weighted variance of the best bid and ask prices (simply the ``best'' price $B(t)$ at time $t$ as the behavior is the same for both sides of the queue, see \cref{fig:varianceBestAskFigure}) appeared to monotonously decrease with increasing $d$ and, to increase exponentially with increasing $f_{\text{m}}$, coming to a negligible distance from $0$ for sufficiently large values of the delay $d$.

\subsubsection{Shape of the average impact scenario.}
Turning our attention to the scenario of an impactful trade occurring at time $t_{\textrm{I}}$, we define the climb $C$ to be the immediate increase in the best price $B$ following a large ($10$-$100$x the size of average market order) trade against the respective order queue. We further define $F$ to be the difference between the highest and the lowest point the best price attains after $t_{\textrm{I}}$, and $I$ to be the long term impact of the trade, i.e.\ the difference between the equilibrium best price prior to the impactful trade and the equilibrium price to which the best price ``settles'' long after $t_{\textrm{I}}$.
We expressed the volume of the large trade considered as a fraction of the volume available on the respective order queue at the time the trade is executed and denote it by $g$. 

\begin{figure}
    \centering\includegraphics[scale=1]{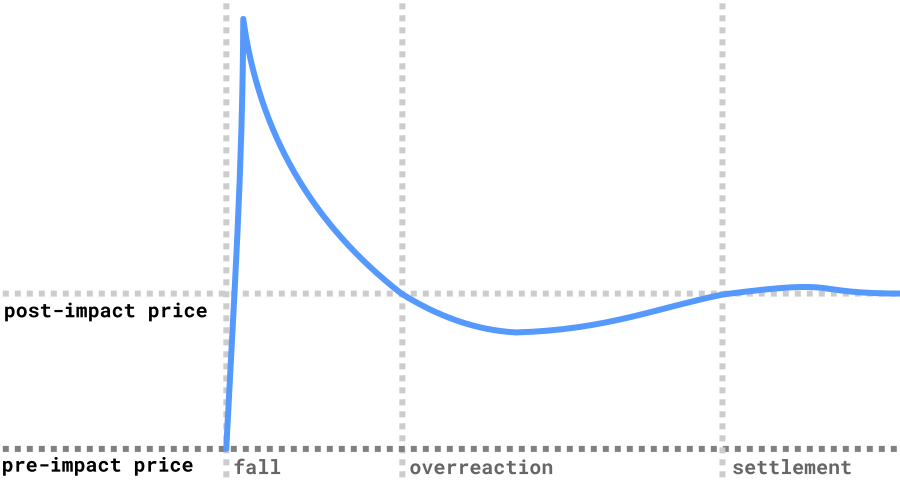}
    \caption{Shown is the shape of the average best price evolution after suffering a large aggressing trade.}
    \label{fig:priceImpact}
\end{figure}

We have found empirically that, irrespective of the volume of the large trade affecting the best price, $e[B](t)$ seemed to exhibit the same feature of going through the phases of \emph{fall}, \emph{overreaction}, and \emph{settlement} (see \cref{fig:priceImpact}).
The climb in the best price itself occurred almost instantly after $t_{\textrm{I}}$ in the vast majority of cases, with the exception when a delayed limit order unaware of the sudden price movement significantly improved the new best price but was quickly eliminated by newly incoming marketable orders. The first phase, fall, exhibited a steep best price fall towards the future equilibrium and its steepness decreased with increasing latency $d$. The fall was succeeded by something that could described as an overreaction, a phase during which the best price dived further below the future equilibrium price and hit the absolute minimum at the time at which the bid-ask spread was also minimal. The best bid and ask prices then diverged again towards their new equilibrium in the settlement phase.

The identification of such patterns has the potential of being of practical utility. They might endow us with a method for predicting the price at which the best price will settle after a large trade given the information about the long-term variance of and current information about the values of the bid-ask spread. 

\subsubsection{The large trade scenario.}
\begin{figure*}
    \centering

    \captionsetup[subfigure]{justification=centering}
    \begin{subfigure}[t]{.3\linewidth}
        \centering\hspace{-0.8cm}\includegraphics[width=1.0\linewidth, height=1.0\linewidth]{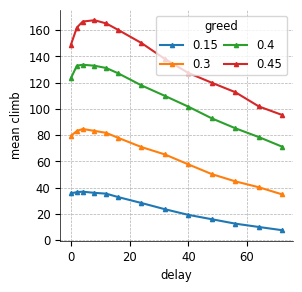}
        \caption{}
        \label{fig:meanClimbFigure}
    \end{subfigure}
    \begin{subfigure}[t]{.3\linewidth}
        \centering\hspace{-0.85cm}\includegraphics[width=1.0\linewidth, height=1.0\linewidth]{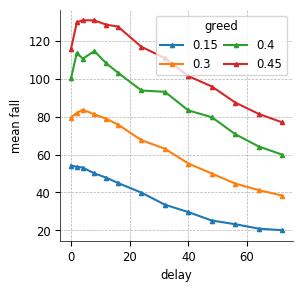}
        \caption{}
        \label{fig:meanFallFigure}
    \end{subfigure}
    \begin{subfigure}[t]{.3\linewidth}
        \centering\hspace{-0.85cm}\includegraphics[width=1.0\linewidth, height=1.0\linewidth]{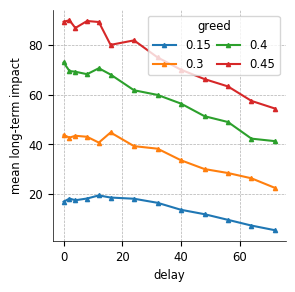}
        \caption{}
        \label{fig:meanLongTermImpactFigure}
    \end{subfigure}
\caption{Absolute mean climb, mean fall, and mean long-term impact (in price ticks) plotted against the processing delay $d$ (in multiples of $t_\text{p}$) for fixed values of $f_\text{m}$ and varying values of greed $g$.}
\end{figure*}

We observed that both $e[C]$ and $e[F]$ decreased linearly for large delays and small delays with small values of $g$ (\cref{fig:meanClimbFigure} and \cref{fig:meanFallFigure}). In addition, large values of $g$ seemed to allow the climb and fall to peak at a specific small delay.

The long-term impact appeared to be mostly linear with $d$ with the downwards slope decreasing with the increasing values of greed, increases linearly with $f_{\text{m}}$ (\cref{fig:meanLongTermImpactFigure}). Furthermore, it did not seem to exhibit the same peak as climb and fall do, demonstrating that these two compensated for each other in the long run. Furthermore, the logarithm of the long term impact increased proportionally to the volume traded, in keeping with the results presented in \cite{potters2003more}.

We said that the best price had reached \emph{stability} if the moving average with a fixed window of size $w$ had fallen within the distance of $\sqrt{\frac{v[B]}{w}}$.

\begin{figure*}
    \centering
    \medskip

    \captionsetup[subfigure]{justification=centering}
    \begin{subfigure}[t]{.3\linewidth}
        \centering\hspace{-0.8cm}\includegraphics[width=1.0\linewidth, height=1.0\linewidth]{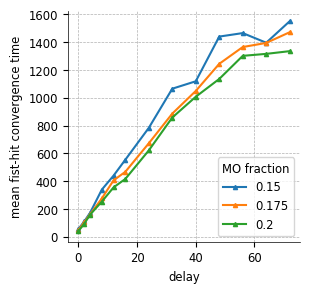}
        \caption{}
        \label{fig:meanFirstHitConvergenceTimeFigure}
    \end{subfigure}
    \begin{subfigure}[t]{.3\linewidth}
        \centering\hspace{-0.85cm}\includegraphics[width=1.0\linewidth, height=1.0\linewidth]{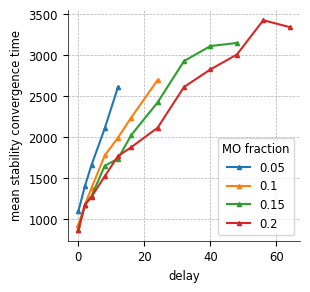}
        \caption{}
        \label{fig:meanStabilityConvergenceTimeFigure}
    \end{subfigure}
    \begin{subfigure}[t]{.3\linewidth}
        \centering\hspace{-0.85cm}\includegraphics[width=1.0\linewidth, height=1.0\linewidth]{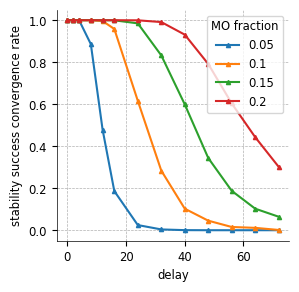}
        \caption{}
        \label{fig:stabilityConvergenceCountFigure}
    \end{subfigure}
\caption{Convergence statistics shown against $d$ (as a multiple of $t_\text{p}$) for different values of $f_{\text{m}}$. The time is also expressed as a multiple of the mean order placement rate $r_{\textrm{p}}.$}
\end{figure*}

Whilst we found significant evidence that the impact of a large trade on the best price depends on the greed parameters, perhaps surprisingly, the mean and variance of the time did not seem to exhibit any notable dependence on the level of greed, i.e.\ the best price appears to converge to stability in time independent of the size of the large trade nor the share of the marketable orders $f_{\text{m}}$ (\cref{fig:meanStabilityConvergenceTimeFigure}). 

Further evidence of such behavior was found when producing the results depicted in \cref{fig:stabilityConvergenceCountFigure}. Here, we looked at the proportion of the runs of the simulation in which the price fulfilled the post-impact stability criterion given above before the simulation was terminated. As can be seen from the plot, simulation runs for higher values of the parameter $f_m$ would see the price succeed to become stabilised in the time horizon of the simulation more often than for the lower values, but the greed parameter had again little to no effect on the proportion of the runs that would become stabilised for varying values of $d$. This is further supported by setting a time limit on convergence in the distant future from the impactful trade and measuring the convergence success rate, defined as the proportion of the simulation runs that succeeded in converging before that time (see \cref{fig:stabilityConvergenceCountFigure}).

\section{Example use case-- market-based coordination of learning agents} \label{sec:example_use}

The generality of MAXE's architecture allows us to simulate multi-agent systems with no relation to finance at all. As a specific example we now implement the experiment presented in \cite{calliess2008no}, concerning a multi-agent mechanism that, under some assumptions about agents' rationality, gives plausible solutions for routing problems.

Consider a finite directed graph $G = (V, E)$, in which each edge $e \in E$ has capacity $\gamma(e)$ and a fixed intrinsic cost $c^e$ for each unit of flow that is to be directed through the edge.
Suppose that there are players who each want to send an amount of flow $d_{sr}$ from some vertex $s$ to $r$ -- that is, player $P_{sr}$ wants to direct $d_{sr}$ from $s$ to $r$ and to that end has an individual plan represented by a vector $\mathbf{f}_{sr} = (f_{sr}^e)_{e \in E}$.
We require the players to plan in such a way that the resulting flows through the graph conserve flow.
The players' planning is influenced by two soft constraints: $\beta_e(\nu) = \max\{0, u_e\nu\}$ where $\nu = f_{sr}^e - \gamma(e)$ is the amount by which the flow directed through $e$ exceeds $e$'s capacity, and $\beta_{sr}(\nu) = \max\{0, u_{sr}\nu\}$ where the argument $\nu = -d_{sr} + \sum_{e \in E}f_{sr}^e$ is the amount by which the player's plan exceeds her demand.

On the implementation side we shall represent every player by an agent. Then, following the approach of \cite{calliess2008no} we transform the soft constraints of the problem by introducing two additional groups of adversarial agents: one for edge, and one for demand constraints. These agents selfishly choose prices for exceeding the edge capacities and failing to meet players' demands, respectively.

In particular, at the beginning of every iteration, each (player-, edge-, demand-) agent decides on their plan following the Greedy Projection Algorithm \cite{zinkevich2003online}. Player-agents decide on how to direct the flow, whereas the adversarial edge- and demand-players decide on the price they are going to charge the player-agents for their respective constraint violations. Player-agents then poll the adversarial agents on their prices and store the information for the decision-making in next iteration of the game.

\begin{figure}
    \centering\includegraphics[scale=0.5]{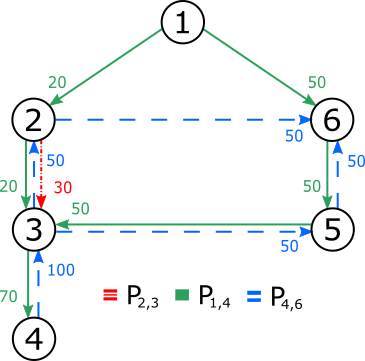}
    \caption{A graphical representation of the experiment output consistent with the simulation run in \cite{calliess2008no}.}
    \label{fig:networkExperiment}
\end{figure}

A simple example of an averaged player plan after a number of iterations is depicted in \ref{fig:networkExperiment}. In this experiment, following \cite{calliess2008no} to the letter, we had a 6-node network and three players $P_{2,3}$, $P_{1,4}$, and $P_{4,6}$ with demands $30$, $70$ and $110$. We set $c^{(2,3)},c^{(3,2)} = 10$, and $c^e = 1$ for all other edges $e$. Edges $(5, 6)$ and $(6, 5)$ had capacities of $50$, while all other capacities were $100$.

Figure \ref{fig:networkExperiment} also shows the output (the resulting flows) of our simulation run that ended up being entirely consistent with the solution found in \cite{calliess2008no}.

\section{A performance comparison} \label{sec:performance_comparison}
Although numerous software packages can be used for simulation of financial markets, we identified ABIDES as a top contender for MAXE, being extensible enough to allow for almost arbitrary simulation of limit order books while featuring a small group of default agent types allowing for an easy simulation setup.

MAXE and ABIDES vary fundamentally in how they approach simulation. While MAXE uses its own, general messaging protocol, ABIDES uses a combination of NASDAQ ITCH-OUCH protocols and agent wake-up scheduling.
To examine how these two different approaches affect the simulation performance we considered one of the simplest multi-agent market models conceivable, in which agents require only very little computation to decide how to act.

Inspired by ABIDES' RMSC01 configuration, we thus started with a unit population for ABIDES consisting of a single market-maker and $25$ ABIDES-default zero-intelligence agents, and a population for MAXE of the same size consisting of MAXE's equivalent agents.
We scaled the unit population by the factors of $1,2,4,8,16$, while examining the average runtime of $1$ hour of simulation time over $10$ attempts for factors $1,2,4,8$. In the case of the $16$-factor, only three runs were considered due to the large demands on memory and duration of the program run.

\begin{figure}
    \centering\includegraphics[scale=0.6]{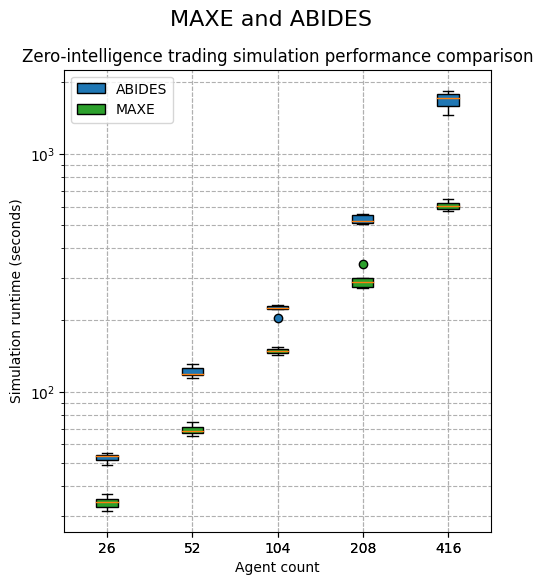}
    \caption{A runtime comparison of MAXE and ABIDES.}
    \label{fig:runVsAbides}
\end{figure}

The results of our simulation are shown in Figure \ref{fig:runVsAbides}.
We only comment on the relative runtime performance of ABIDES and MAXE as the agent population increases.

For small agent populations there does not seem to be a significant difference in the performance of the two simulators.
As the agent population grows, the runtime of MAXE becomes more clearly separated from that of ABIDES.
This is most likely due to the different methods of agent communication handling employed by the two simulators.
Our plot stops at 416 agents, after which the memory demand of the Python environment running ABIDES exceeded the 16GiB of memory available in our small workstation and the operating system resorted to swapping, significantly hindering the runtime performance of ABIDES.
A test run of ABIDES on 450 agents resulted in a Python \textit{Memory error}.
We noticed that ABIDES, at present, does not allow for regular flushing of trading history to the disk, and we believe that a small adjustment to the design of the simulator, coupled with the employment of an appropriate memory management strategy, could resolve the memory greediness currently limiting the feasibility of ABIDES' use when a larger number of agents is involved. For comparison, a simulation consisting of a population of 100,022 agents in MAXE with the same setup fit comfortably into 100MiB of memory.

In summary, the plot of Figure \ref{fig:runVsAbides} is consistent with our intuition that the design choices made for the key components of MAXE do indeed result in marked performance improvements over alternative packages when a larger population of agents is to be simulated.

\section{Conclusions}\label {sec:conclusions}
We have introduced a new multi-agent simulation framework for financial market microstructure, called the Multi-Agent eXchange Environment (MAXE). There are a number of distinctive advantages MAXE offers over alternative simulation frameworks such as ABIDES \cite{byrd2019abides}. Most notably, our framework was designed to be fast and flexible; it allows the modelling of different matching rules and can model latency.  

We have demonstrated its potency for research into market dynamics. In particular, we utilised MAXE to showcase a mini study of the impact the delay in processing order has on a few LOB statistics and on the behaviour of the best prices after a large trade is registered with the exchange. We have also shown that MAXE is suitable for applications beyond the study of financial markets, as we used it to simulate a multi-agent reinforcement-learning network routing scenario.
As our first evidence provided suggests, MAXE can be used to simulate markets and multi-agent systems more efficiently than comparable existing toolboxes.
We therefore hope that it will be useful to facilitate research across different disciplines in need of  simulating large-scale agent-based models.  


Expanding on our illustrative case studies would be interesting in particular, given the dearth of studies utilising ABM in the context of pro-rata matching rules. We hope such inquiries would be greatly aided by our MAXE package, providing a standardised, scalable, and easily customisable toolbox to support this kind of research. 


%
%
%
%
\bibliographystyle{splncs04}
\bibliography{prorata.bib}

\end{document}